\documentclass[12pt]{article}

\begin{document}

\baselineskip=18.6pt plus 0.2pt minus 0.1pt

 \def\be{\begin{equation}}
  \def\ee{\end{equation}}
  \def\bea{\begin{eqnarray}}
  \def\eea{\end{eqnarray}}
  \def\nn{\nonumber\\ }
\newcommand{\nc}{\newcommand}
\nc{\bib}{\bibitem}
\nc{\cp}{\C{\bf P}}
\nc{\la}{\lambda}
\nc{\C}{\mbox{\hspace{1.24mm}\rule{0.2mm}{2.5mm}\hspace{-2.7mm} C}}
\nc{\R}{\mbox{\hspace{.04mm}\rule{0.2mm}{2.8mm}\hspace{-1.5mm} R}}

\begin{titlepage}
\title{
{\bf  Relation Between Holonomy}\\[.3cm]
{\bf Groups in Superstrings, M and F-theories}
\author{Adil Belhaj$^{1,2,3}$\thanks{\tt{belhaj@uniar.es}; GNPHE/0808}, Luis  J. Boya$^2$\thanks{\tt{luisjo@uniar.es}},
Antonio Segui$^2$\thanks{\tt{segui@uniar.es}}\\
{ \small $^1$ Centre National de l'Energie, des Sciences et des Techniques Nucl\'eaires,  CNESTEN} \\
{\small Rabat, Morocco} \\{ \small $^2$ Departamento de F\'{\i}sica
Te\'orica,
Universidad de  Zaragoza} \\
{ \small E-50009-Zaragoza, Spain} 
\\
{\small $^3$  GNPHE, Point  focal:  Rabat, Morocco}
\\
 } } \maketitle
\thispagestyle{empty}
\begin{abstract}
We  consider   manifolds with special holonomy groups $SU(3)$, $G_2$ and $Spin(7)$
as suitable for  compactification  of superstrings, M-theory
and F-theory (with only one time)  respectively.    The  relations  of these groups with
 the octonions are discussed,  reinforcing their  role  in  the physics of string theory
  and duality. We also  exhibit three triple exact sequences   explaining the
 connections between the mentioned  special holonomy groups.
\end{abstract}
{\tt  KEYWORDS}:  Superstrings, M-theory, F-theory, Duality, Octonions,  Holonomy Groups.
\end{titlepage}

\newpage
\tableofcontents
\newpage
\section{Introduction}
 Once  one includes strings and/or other extended objects, extra dimensions  became  unavoidable,
 and thence the  necessity  of dimensional  reduction, introducing tiny compactifying  spaces, as
  we see only four extended dimensions. Over the few past years, there has been an increasing
   interest in studying duality in superstring theory  and supersymmetric  models, related to compactification.
   Interesting examples are mirror symmetry  in four dimensions between pairs of Calabi-Yau
threefolds in type II superstrings \cite{vafa} and  their strong/weak coupling dualities
    with  heterotic superstrings on $K3\times T^2$  \cite{KV}.
 The most important consequence
of the study of string duality is that all (five) superstring models are equivalent in the sense that
they correspond to different limits in the moduli space of the same theory, called M-theory  by Witten
\cite{witten,townsend,PT,adil1,adil2}. M-theory   is considered nowadays as the best candidate for the
unification
of the weak and strong coupling sectors of superstring models, and is described, at low energies,
 by eleven dimensional supergravity theory. In this paper, we shall also consider F-theory
 in 12 dimensions, initiated  by C. Vafa \cite{vafaft},  with metric of signature (2,10)
  (the original idea)  but also (1,11).

To preserve adequate supersymmetry (Susy) in our  $4$d  these intermediate spaces have to
 behave as special holonomy manifolds. In particular, to obtain ${\cal N}=1$ supersymmetry down
  to four dimensions, which is necessary to hold a phenomenologically  acceptable  chiral
    theory, we need a manifold with $G_2$ holonomy if coming from 11d M-theory
    \cite{witten,townsend,PT,Duff,AW,adil1,adil2,adil3,adil4}, or  a Calabi-Yau three-fold
     with SU(3)  holonomy if starting from heterotic superstrings  \cite{witten1}.
 The first case is due to the fact that the $G_2$  group is the maximal
  subgroup of $SO(7)$, for which the
eight dimensional spinor representation of $SO(7)$ can be decomposed as the fundamental representation
of $G_2$ and one singlet.  In the second case, as $Spin(6)=SU(4)$, the  $SU(3)$
subgroup shows again  the $4=1+3$ reducible  representation, guaranteeing ${\cal N}=1$  down to $4d$;
 the further break, e.g,  from one factor of  $E_8 \times  E_8$  to $E_6 \times SU(3)$  is well studied for example in
   \cite{witten1}.  Now the conventional F-theory of  Vafa \cite{vafaft} has signature $(2,10)$;
    one can consider however an F-theory with conventional metric $(1,11)$.  In this case, the holonomy
     of the compactifying  manifold might be $Spin(7)$, the largest of the exceptional holonomy
    groups. The alternative case  is $SU(4)$  corresponding to Calabi-Yau four-folds.
     Very recently, a GUT realization has  been given in the context of F-theory for such
     manifolds with elliptic K3 fibration
      \cite{BHV1,vafa08}.

As a whole,  there are many ways to get four dimensional models using different compactifications as intermediates.
These constructions could be related to each other  due to  different
  sort of dualities, which appear in the process. As an example, we point out here several
   equivalences in seven dimensions giving rise to a web of dualities (with F and M for F-theory,
   M-theory, \ldots) \cite{vafa}
\begin{equation}
F/K3 \times S^1\sim het/T^3\sim  IIB/S^2\times S^1\sim M/K3\sim IIA/S^2\times S^1.
\end{equation}
This   relation of special  holonomy manifolds and dualities  can be pursued, of course,
 to lower dimensions. In particular, one might naturally ask  for similar     relations
 in  four dimensions involving the special holonomy manifold compactifications previously
 mentioned. In this work, we  address  the problem of dualities  and semi-realistic compactifications,
 as regards the different holonomy groups.  We shall focus mainly in three groups, $SU(3)$,
 $G_2$ and $Spin(7)$ suitable for superstrings, M-theory and F-theory. We shall not be so
  much concerned with the manifolds themselves \cite{Joyce}. The relation of the above groups  with
   octonions should   be apparent; we shall devote some space to study it. We shall
    exhibit also some exact triple sequences, which we believe illuminate the relations between these groups and several subgroups.

The organization of this paper   is as follows. In section 2 we recall the classification
 of special holonomy manifolds (Berger 1955). In section 3, we review different ways of
  constructing    four dimensional models with minimal number of supercharges  from  higher dimensional
   supersymmetric theories.  The exact sequences mentioned above are explained in section 4.
       Background on toroidal  M-theory compactification,  division algebras and related group manifolds
        are  included in the two appendices.
\section{Special holonomy manifolds}
\subsection{Generalities on holonomy groups}
Extended objects, unification of forces and supersymmetry, all suggest extra dimensions
 of space-time. However we see only   $4= (1+3) $ dimensions: Some mechanism has to be advocated to
 prevent  the extra size  of the space to be visible. The compactification is the most
 accepted ingredient, namely making the extra  dimensions too small to be observable. In the original
 Kaluza-Klein type of theories, the observable  gauge group  in 4 dimensions came from
the {\it isometry} groups of the compactifying space (this is why
the U(1)  gauge group
  of electromagnetism came from circle compactification).  But when supersymmetry is present,
it was realized in the early 80s  that  the {\it holonomy} groups of
intermediate spaces  respond of the number of supercharges surviving
in four dimensions \cite{awada83}.

Here  we recall briefly the classification of  special holonomy groups and manifolds
 of Berger (1955) in a form suitable for all later physical
 consideration\footnote{ Note that
the book of Joyce \cite{Joyce} is the best modern reference  for
this subject.}.
 Let  $\cal M$ be any  $n$ dimensional differentiable manifold. The structure group of the tangent bundle
is a subgroup of the general linear group, $GL(n,R)$. Now the
maximal compact group of the linear group is $O(n)$. So the quotient
homogenous space $ \frac{GL(n,R)} {O(n)}$ is a contractible space;
hence, a manifold admits always a Riemannian metric $g$, whose  tangent
structure  group  is (a subgroup of) the orthogonal group.  The isometry
group $Iso(\cal M)$  is the set of diffeomorphisms  $\sigma$ leaving
$g$ invariant. For spheres we have $Iso(S^n)=O(n+1)$; for torii  $Iso(T^n)=U(1)^n$.

For an arbitrary $n$ dimensional Riemannian manifold $\cal M$, the
structure group of the tangent  bundle is, as said, a subgroup of
$O(n)$. Carrying a orthonormal frame $\epsilon$ of $n$ vectors in a
point $P$ through  a closed loop $\gamma$ in the manifold
\begin{equation}
\gamma: P \to P'\to P
\end{equation}
it becomes another frame $\epsilon'=o \cdot \epsilon $  which is  shifted by certain  element $o$ of $O(n)$.
This is the {\it holonomy element} of the loop. All elements of all possible loops on the manifold
 from $P$ make up
the holonomy group of the  manifold $Hol_P(\cal M)$,  which is  always a subgroup of $O(n)$, and is easily
 seen to be independent of the point $P$ for an arcwise-connected manifold.
 A generic Riemannian manifold  would have holonomy $O(n)$, or $SO(n)$ if it is orientable, whereas the isometry
 group generically is just the identity; in a way isometry and holonomy are complementary.

For any vector bundle with connection, the structure group reduces to the holonomy group (reduction theorem).
The corresponding Lie algebra of the holonomy group is generated by the curvature of the connection
(the Ambrose-Singer theorem)\cite{KN63}.

 \subsection{ Classification of special holonomy manifolds}
Only special groups can act as holonomy groups; the classification
of possible holonomy groups was  carried by M. Berger in 1955. If
the manifold is irreducible, $ Hol ({\cal M})$ should be in $ O(n)$.
Its Lie algebra, as we said, is generated by the curvature. For the
irreducible non symmetric case, there are three double series
 corresponding to the numbers $\mathbf{R},\mathbf{C}$ and $\mathbf{H}$, and two isolated cases
  related to the octonion numbers $\mathbf{O}$.  For  each number domain there are the  generic
 case and the unimodular subgroup restriction.  The classification is given in the following Table \cite{Boya1}
\begin{center}
\begin{tabular}{lll}
Numbers & Group & Unimodular Form \\
\hline \\
$\mathbf{R}$ & $O(n)$ & $SO(n)$ \\
& generic case & orientable, $w_{1}=0$ \\
$\mathbf{C}
$ & $U(n)$ & $SU(n)$ \\
& K\"{a}hler, $d\omega =0$ & Calabi-Yau, $c_{1}=0$ \\
$
{\mathbf{H}}
$ & $Sp(1) \times/_2 Sp(n)$ & $Sp(n)$ \\
& Quaternionic &  Hyperk\"{a}hler \\
$\mathbf{O}$ & $Spin(7)$ in $8d$ spaces & $G_{2}$ in $7d$ spaces \\
& $Oct(1)$ & Aut$\mathbf{O}$%
\end{tabular}
\end{center}

We provide some explanations.  An arbitrary $n$-dimensional Riemannian manifold
$ \cal{M}$  has $O(n)$  as the maximal holonomy group. The obstruction
   to   orientability is measured  by the first  Stiefel-Witney class of the tangent bundle,
   $w_1({\cal M})=w_1(T{\cal M}) \in H^1( {\cal M} ,Z_2)$.\\
A  $n$-dimensional complex K\"{a}hler  manifold parameterized  by $z_i, i=1,...,n$  has
 a closed  regular real K\"{a}hler  two form  $\omega$ given in a local chart by
\begin{equation}
\omega=iw_{i{\bar j}}dz_i\wedge d{\bar z_{\bar j}}, \quad d\omega=0
\end{equation}
and its holonomy group is $U(n)$.  Now as $ \frac{U(n)}{SU(n)}=U(1)$, we have the diagram

\begin{equation}
\begin{tabular}{lllll}
$\;$$\;$SU(n) &  &  &  &  \\
$\quad$ $\;$ $\downarrow $ &  &  &  &  \\
$\;$ U(n) & $\longrightarrow $ & B & $\longrightarrow $ & M \\
det $\downarrow $ &  &  &  &  \\
$\;$$\;$U(1) & $\longrightarrow $ & B%
\'{}
& $\longrightarrow $ & M%
\end{tabular}%
\end{equation}
where   the middle line  is the   frame bundle: $B$ is  the principal bundle of orthogonal unitary frames.
 The last bundle  is mapped  to an element of $H^2({\cal M}, Z)$; hence, the determinant map defines the
  first Chern class of the bundle as $c_1({\cal M}) \in H^2({\cal M},Z)$.
It turns out that when $c_1=0$, the K\"{a}hler manifold becomes a Calabi-Yau manifold  with
$SU(n)$  holonomy group and it is Ricci flat, $Ric=0$. Note that one-dimensional Calabi-Yau manifold  is
nothing but a complex elliptic curve. Then its Hodge diagram is given by
\def\m#1{\makebox[10pt]{$#1$}}
\begin{equation}
  {\arraycolsep=2pt
  \begin{array}{*{5}{c}}
    &&\m{h^{0,0}}&& \\ &\m{h^{1,0}}&&\m{h^{0,1}}& \\
    &&\m{h^{1,1}}&&
  \end{array}} \;=\;
  {\arraycolsep=2pt
  \begin{array}{*{5}{c}}
    &&\m1&& \\ &\m1&&\m1& \\ &&\m{1}&&
  \end{array}}
\end{equation}
The second example of   such geometries  is the  K3 surface with  $SU(2)$ as holonomy group.
Its  Hodge diagram reads \cite{Boyafest} as
\def\m#1{\makebox[10pt]{$#1$}}
\begin{equation}
  {\arraycolsep=2pt
  \begin{array}{*{5}{c}}
    &&\m{h^{0,0}}&& \\ &\m{h^{1,0}}&&\m{h^{0,1}}& \\
    \m{h^{2,0}}&&\m{h^{1,1}}&&\m{h^{0,2}} \\
    &\m{h^{2,1}}&&\m{h^{1,2}}& \\ &&\m{h^{2,2}}&&
  \end{array}} \;=\;
  {\arraycolsep=2pt
  \begin{array}{*{5}{c}}
    &&\m1&& \\ &\m0&&\m0& \\ \m1&&\m{20}&&\m{1.} \\
    &\m0&&\m0& \\ &&\m1&&
  \end{array}}
\end{equation}
Since $SU(2)=Sp(1)$, the K3  surface is also a hyperk\"{a}hler (Calabi-Yau) manifold.
 Notice that Hyperk\"{a}hler manifolds are also Calabi-Yau, but the quaternionic
 manifolds in general are not. Quaternionic manifolds have for holonomy
 $Sp(1) \times Sp(n)/Z_2$, abreviated as $Sp(1) \times/_2 Sp(n)$ in the Table.

Finally, the  two cases related to the octonions are  $G_2=Aut (Octonions)$  and $Spin(7)$.
The former is well known  and we shall elaborate on it later; as for the $"Oct(1)"$
label for the later, this  will also be justified in the Appendix $2$.

Note that, in general,   a manifold  with a specific holonomy  group $ Hol( {\cal M})=G$
 implies  the manifold carries an additional structure, preserved by the group $G$. For
 example, an orientable  manifold, with holonomy  within $SO(n)$,   has an invariant volume element, indeed
$SO(n)=O(n)\bigcap SL(n,R)$.  A  K\"{a}hler  manifold, with holonomy inside $U(n)$
has an invariant closed 2-form as $U(n)=O(2n)\bigcap Sp(n)$. A manifold with $G_2$ holonomy
 will carry an invariant $3$-form, etc.
\section{  Semi-realistic compactifications}
Consider first superstring  theory, which lives in ten dimensions
\cite{Schwarz}; down to four dimensions we want only ${\cal N} =1$,
i.e. four supercharges, as to allow for parity violation. We know
that compactification on  a $SU(n)$-holonomy manifold would reduce
the supercharges by a factor of $1/2^{n-1}$, so $SU(3)$-holonomy
(1.e., a Calabi-Yau $3$-fold) would be just right to descend from
the heterotic string ($16$ supercharges) to a four  dimensional model with only four superchages.  Indeed, the search for
CY$_3$ manifolds was a big industry in the $80$s \cite{Candelas}.
This choice is also natural, as  $SU(3) \subset SU(4)=Spin(6)
\rightarrow SO(6)$, and obviously $4=3+1$ leaves one surviving
spinor.

In M-Theory with $11$ dimensions the preferable compactifying manifold would be one with $G_2$ holonomy:
now the inclusions are $G_2 \subset SO(7) \leftarrow Spin(7)$, and $8=1+7$, as $2^{(7-1)/2}=8$, type real.
$G_2$-holonomy manifolds are also Ricci flat; they were first proposed for the M-Teory in \cite{PT}.
Indeed, $G_2$ holonomy manifolds would preserve $1/2^3$ supercharges, and in $11$d there are $2^{(11-1)/2}=32$,
type real again as $10-1=9 \equiv 1$ mod $8$.

We can also consider $8$d compactifying manifolds in at least two context: $1$) Descend $11 \rightarrow 3$ just
for illustrative purposes, and $2$) F-Theory with metric ($1,11$); the original sugestion of Vafa was $12=(2,10)$,
see \cite{vafaft, Boya2}. Here the manifolds of choice would be either CY$_4$, that is, $SU(4)$-holonomy manifolds,
preserving $4$ charges out of $32$ (which is what we want), or $Spin(7)$, the last of the exceptional
holonomy groups; $Spin(7)$ does the job as it has an irreducible $8$d representation, same as $Spin(8)$ and
$Spin(7) \subset Spin(8)$.

Finally, starting from the conventional F-Theory with signature $(2,10)$ it is necessary to compactify in
 a manifold with signature $(1,6)$. The following Table sums up the situation:

 \begin{center}
\begin{tabular}{lll}
Theory & \quad Dim Change &  \qquad Holonomy \\
\hline \\
Heter. String & $\quad 10d\longrightarrow 4d$ & $\qquad SU(3)$ (CY$_3$ manifold) \\
M-theory  & $\quad 11d\longrightarrow 4d$ & $\qquad G_2$ (Ricci flat) \\
M-theory  & $\quad 11d\longrightarrow 3d$ & $\qquad Spin(7)$ (Ricci flat) \\
F-theory $(1,11)$  & $\quad 12d\longrightarrow 4d$ & $\qquad Spin(7), SU(4) ($CY$_4)$ \\
F-theory $(2,10)$ & $\quad 12d\longrightarrow 4d$ & \qquad Indefinite form of $Spin(7)$%
\end{tabular}
\end{center}
\bigskip

Let us  comment this table. There is clearly a double  inclusion
\begin{equation}
SU(3) \mbox{(dim 8, rep. 6)} \subset G_2 \mbox{(dim 14, rep. 7)} \subset Spin(7) \mbox{(dim 21, rep. 8)}
\end{equation}
linked with the dimensions $(10,11$ and $12)$ of the physical (!) theories.
It is remarkable  that these three groups have a neat relation with octonions, which we shall
elaborate now. Also, the related question arises: given the many duality relations extant in string
 (and M) theories,
is there any connection between the different holonomy groups? We shall take up this question
 in the next section $4$.

\subsection{ Relation with octonions}
The necessary properties we need of the division algebra of the octonions $\mathbf {O}$ are
 described in the Appendix $2$. Here we recall that $G_2$ is the automorphism group of
  the octonions (as $SO(3)$ is $Aut(\mathbf{H})$ and $Z_2=Aut(\mathbf{C})$ ); the reals
   $ \mathbf {R}$ have not autos, hence the representation $8$ of $G_2$ in the octonions
    split naturally in $8=1+7$. Note that  $G_2$ acts transitively in the $S^6$ sphere of unit imaginary octonions.
     This implies the $6$-sphere acquires a quasi-complex structure (Borel-Serre). The
     sequence  reads as follows
\begin{equation}
\label{su3}
SU(3) \rightarrow  G_2 \rightarrow S^6 \quad (8+6=14)
\end{equation}

Now the octonionic product preserved by $G_2$, as any
 algebra $( xy=z)$, defines  an invariant  $T^1_2$  tensor  and  the conservation of the norm is like
 preserving a quadratic form.  The $T^1_2$ tensor can be seen  then as a $T^0_3$  tensor. Now the
  alternating property of the octonionic product is equivalent to this $T^0_3$ tensor to  become a 3-form in $R^7$, $ \wedge T^0_3$, which is generic, (i.e., they make up an open set). This implies
\begin{equation}
\mbox{dim}\; G_2= \mbox{dim}\; GL(7,R)-\mbox{dim} \;\wedge T^0_3=49-35=14.
\end{equation}
Besides, the dual form $\wedge T^0_4$ is also invariant, implying  $G_2$ is unimodular, i.e.
lies inside $SO(7)$. The dimension $14$ of this $G_2$ can of course be proved directly \cite{Rosenfeld}.

As with respect to $Spin(7)$, it has a real $8$-dimensional
representation as we said, and hence it acts in $S^7$, indeed
transitively. The little group acts in the $S^6$ equator, and it
 is certainly $G_2$. In fact, there is some sense, as explained in the Appendix $2$, to call $G_2$ and $Spin(7)$
  respectively $SOct(1)$ and $Oct(1)$.

\section{Relation between holonomy groups}
We know that strings,  M and F theories are related by different sorts of dualities and dimensional
 reductions. As a consequence we expect that also the different holonomy groups used in the different
  compactifications should be connected.
The aim of this section is to adress this question  using   exact sequences and commutative diagrams
for these groups.  To start,  we  note the following. If $H \subset G$  with (left)-coset space $X$,
 we write $H \to G  \to  X$ for  $ G/H =  X$; when $H$ is normal, $X$ becomes the quotient group.
   Some of these diagrams   have been already  given  in \cite{Boya3}.\\
The  first diagram   that we present here  comes form the inclusion  of the exceptional holonomy group
$ G_2 \subset Spin(7)$. The later acts transitively in all units in $\mathbf{O}$ (octonions of norm one
 and octonions with imaginary part of norm one), whereas $G_2= Aut(\mathbf{O})$  obviously leaves $1$
  invariant (the real part of the octonion).
 So  the main cross of the diagram  takes the following form 
\bigskip

\begin{equation}
\begin{tabular}{lllll}
&  & Spin(6) &  &  \\
&  & $\quad \downarrow $ &  &  \\
$G_{2}$ & $\longrightarrow $ & Spin(7) & $\longrightarrow $ & $S^{7}$ \\
&  & $\quad \downarrow $ &  &  \\
&  & $\quad S^{6}$ &  &
\end{tabular}%
\end{equation}
where the vertical line is elemental\footnote{The spin groups acting  on  the natural spheres via the $SO$
 (covered) groups.}. With the $A_3=D_3$ isomorphism $Spin(6)=SU(4)$  and the fact that
  $ SU(3)\subset G_2\bigcap (Spin(6) =SU(4))$, we can complete the previous diagram.  The result is given by
\begin{equation}
\label{seq}
\begin{tabular}{llllll}
$SU(3)$ & $\longrightarrow $ & $SU(4)$ & $=Spin(6)$ & $\longrightarrow $ & $%
S^{7}$ \\
$\quad\downarrow $ &  &  & $\downarrow $ &  & $\parallel $ \\
$\quad G_{2}$ & $\longrightarrow $ &  & $Spin(7)$ & $\longrightarrow $ & $S^{7}$
\\
$\quad\downarrow $ &  &  & $\downarrow $ &  &  \\
$\quad S^{6}$ & === &  & $S^{6}$ &  &
\end{tabular}.
\end{equation}
From this picture   we  can see in particular the octonionic nature
of $SU(3)$. It is a group of
 automorphism of octonions, fixing  the product  $(ij)k$. There is a suspicion, still conjectural,
  that this is the reason why the gauge group of the strong forces  is  $SU(3)$ color.\\
To get  the second diagram,  we start by another  obvious cross, since   $Spin(7)$  is the (universal)
 double cover of  $SO(7)$.  In this way, we have the following diagram
\begin{equation}
\begin{tabular}{lllll}
&  & $Z_{2}$ &  &  \\
&  & $\downarrow $ &  &  \\
$G_{2}$ & $\longrightarrow $ & $Spin(7)$ & $\longrightarrow $ & $S^{7}$ \\
&  & $\downarrow $ &  &  \\
&  & $SO(7)$ &  &
\end{tabular}.
\end{equation}
It is  known  that $G_2$ does not  have a   centre, so $Z_2=Z_2$ must be  the upper row. The rest
 is easy  to complete  since  $S^7/Z_2$ is the real projective space $RP^7$.   We end up   with the
   following  picture
\begin{equation}
\begin{tabular}{lllll}
&  & $Z_{2}$ & $===$ & $Z_{2}$ \\
&  & $\downarrow $ &  & $\downarrow $ \\
$G_{2}$ & $\longrightarrow $ & $Spin(7)$ & $\longrightarrow $ & $S^{7}$ \\
$\parallel $ &  & $\downarrow $ &  & $\downarrow $ \\
$G_{2}$ & $\longrightarrow $ & $SO(7)$ & $\longrightarrow $ & $RP^{7}$%
\end{tabular}.
\end{equation}
From this diagram one can  learn just the lower row, somewhat unexpected, until one sees the middle row.
The lower row is also a remainder that the orthogonal groups have torsion \cite{Borel}.

The  third, final,  diagram is obtained by  asking  the question that $Spin(7)$
 acts transitively and  isometrically in the seventh sphere $S^7$.  Indeed,  it must be a
 subgroup of $SO(8)$. What about the quotient (homogeneous space)?.  To answer this question,
   we   start  first   with the  the following incomplete cross
\begin{equation}
\begin{tabular}{lllll}
&  & $Spin(7)$ & $\longrightarrow $ & $S^{7}$ \\
&  & $\downarrow $ &  & $\parallel $ \\
$SO(7)$ & $\longrightarrow $ & $SO(8)$ & $\longrightarrow $ & $S^{7}$ \\
&  & $\downarrow $ &  &  \\
&  & $??$ &  &
\end{tabular}%
\end{equation}
and then  try to  finish it. Indeed,  $G_2$ lies inside  both  $Spin(7)$  and  $SO(7)$,
 then it must be their intersection and must appear in the upper left corner. The rest
  of  the diagram   can be obtained easily, and the result is
\begin{equation}
\begin{tabular}{lllll}
$G_{2}$ & $\longrightarrow $ & $Spin(7)$ & $\longrightarrow $ & $S^{7}$ \\
$\downarrow $ &  & $\downarrow $ &  & $\parallel $ \\
$SO(7)$ & $\longrightarrow $ & $SO(8)$ & $\longrightarrow $ & $S^{7}$ \\
$\downarrow $ &  & $\downarrow $ &  &  \\
$RP^{7}$ & === & $RP^{7}$ &  &
\end{tabular}.
\end{equation}
The new result we learn is just the middle column.

This completes our study of the holonomy groups which are  suitable for the compactification
 of  superstrings, M, F-theories  respectively.   In particular,   we have studied     their
 relations  with the divisor algebra of octonions. This  may  reinforce   the role of the octonions
  in physics of  strings and higher dimensional  objects  moving  on manifolds with non trivial holonomy groups.  We have also
  found three triple exact sequences   explaining  some  links between  these  groups. One of
  the nice results that one gets from   the  diagrams is that one can also   see   the possible
  connections between the corresponding geometries. Indeed, from the following sequence of inclusions
\begin{equation}
SU(3)\longrightarrow G_2\longrightarrow S^6,
\end{equation}
one can see that the manifold with $G_2$ holonomy  can  be constructed  in terms of Calabi-Yau
 three folds with the $SU(3)$  holonomy group. This has been already discussed in \cite{Joyce}.
  We can suppose the same thing  for the manifold  with  $Spin(7)$  holonomy. It  can be
    constructed either from  manifold with   $G_2$ holonomy  or  Calabi-Yau fourfolds. This
     can be easily seen from the subdiagram (\ref{seq}).

\section{Appendix I: Toroidal Compactification}
 String theory lives in ten dimensions, maximal supergravity and M-theory  in eleven,
  the original F-theory of   Vafa in $12$d.  As we have seen, if we want to get models
  with minimal supercharges, we need compact manifolds with special holonomy groups.

   To complete the study we give here some information about toroidal compactification.
    In particular we consider the case of M-theory.  We start by recalling the particle
     content from 11 dimensional supergravity down to 4 or 3 dimensions.  We suppose a
     step-by-step reduction, so the intermediate manifold  is always a circle. This was
     first found  by Cremmer (ca. 1980), who also showed that scalars make up a sigma model
      type of manifold, as  homogeneous quasi-euclidean spaces. In particular, M-theory compactified
      on $T^k$  has U-duality group $E_k(Z)$ and scalars taking
values in $\frac{E_k}{H_k}$, where $H_k$ is the maximal compact
subgroup of $E_k$.  In eleven dimensions, one has a graviton,
a 3-form gauge field and the gravitino.  We give  the moduli space of the
toridal compactification from this theory; starting in dim $9$, they are given in the
following Table
\begin{equation}
\begin{tabular}{ll}
d=9 & $\frac{SL(2,R)}{SO(2)}\times SO(1,1)$ ($3$ scalars) \\
d=8 & $\frac{SL(3,R)}{SO(3)}\times \frac{SL(2,R)}{SO(2)}$ ($7$ scalars) \\
d=7 & $\frac{SL(5,R)}{SO(5)}$  ($14$ scalars) \\
d=6 & $\frac{SO(5,5,R)}{SO(5)\times SO(5)}$ ($25$ scalars) \\
d=5 & $\frac{E_{6}}{USp(8)}$ ($42$ scalars) \\
d=4 & $\frac{E_{7}}{SU(8)}$ ($70$ scalars) \\
d=3 & $\frac{E_{8}}{SO(16)}$ ($128$ scalars)%
\end{tabular}
\end{equation}

Notice for $d=3,4$ and $5$ the large group is "reconstructed" by a distinguished representation of the subgroup, as
\begin{eqnarray}
&E_8 \sim (adj+spin) \,  \mbox{of} \, O(16) : 120+128= 248 \\
&E_7 \sim (adj+[ 1^4]) \, \mbox{ of} \, SU(8) : 63+70= 133  \nonumber  \\
&E_6 \sim (adj+[ 1^4]') \, \mbox{of} \, Sp(4) : 36+(70-28)= 78 \nonumber
\end{eqnarray}

Beyond this, the descent from $E_6$ has two branches: the $E$-branch $E_6 \rightarrow D_5 \rightarrow A_4$ and
$Sp(4) \rightarrow Sp(2)^2 \rightarrow Sp(2)$ and the octonionic branch $E_8 \rightarrow E_7 \rightarrow E_6
\rightarrow F_4$ with subgroups $O(16) \rightarrow O(12) \times Sp(1) \rightarrow O(10) \times U(1) \rightarrow O(9)$.

\section{Appendix II: The octonions}
We recall here some properties of division algebras in relation with special holonomy
 groups and manifolds. Starting with the real numbers $\mathbf R$, the space $R^2$ becomes
  an algebra $i \equiv \{0,1 \}$ and $i^2=-1$: we get the complex number $\mathbf
  C$.
  It is a commutative and distributive division algebra. Adding a second unit
  $j$, $j^2=-1$ a third $ij$ is necessary, with $ij=-ji$, and we obtain the
   division algebra of quaternions $\mathbf H$ in $R^4$. It is anticommutative
    but still distributive. Adding another independent unit $k$ to $i$ and $j$,
     with $k^2=-1, ik=-ki, jk=-kj$, we have to complete with $e_7=(ij)k$ to the
      algebra of octonions $\mathbf O$ in $R^8$, with units $1,i,j,k;ij,jk,ki;(ij)k=-i(jk)$.
      It is neither commutative nor associative, but still a division algebra:
      if $o=u_0+ \Sigma_{i=1}^7 u_i e_i$ we have
\begin{equation}
\bar o:=u_0 -\Sigma_{i=1}^7 u_i e_i \qquad {\cal N}(o) = \mbox{norm}(o)\! := \bar o o;
\mbox{inverse}\, o^{-1}=\frac {\bar o}{{\cal N}(o)}.
\end{equation}

The associator $\{o_1,o_2,o_3 \}:=(o_1 o_2) o_3-o_1 (o_2 o_3)$ is
 completely antisymmetric. The four algebras
${\mathbf R}{\mathbf C}{\mathbf H}{\mathbf O}$ are composition algebras, that is, we have
 ${\cal N}(xy)={\cal N}(x) {\cal N}(y)$. The continuous
automorphism groups are easily seen to be
\be
\mbox{Aut}({\mathbf R})=1,\quad \mbox{Aut}({\mathbf C})=Z_2, \quad \mbox{Aut}({\mathbf H})=SO(3),
\quad \mbox{Aut}({\mathbf O}):= G_2.
\ee

The norm-one elements form, for ${\mathbf R}{\mathbf C}{\mathbf H}{\mathbf O}$, respectively
\be
O(1)=Z_2=S^0; \quad U(1)= SO(2)=S^1; \quad Sp(1)=SU(2)=Spin(3)=S^3; \quad \mbox {and}\, S^7.
\ee

Now $S^7$ has an invertible product structure, in particular is paralellizable,
 but is not a group, because nonassociativity. Let us name jokingly $'Oct(1)' =S^7$.
 One obtains a {\it bona fide} group by stabilizing $S^7$ by the octonion automorfism
 group $G_2$ \cite{Ramond}. The result is $Spin(7) \approx G_2 (\times S^7$; we shall name $Spin(7):=Oct(1)$.

We recall now the description of compact Lie groups as finitely twisted products
of odd dimensional spheres (Hopf 1941); for details see \cite{Boya4}. For example
in the quaternion case one gets the sequence
\begin{equation}
Sp(1)=Spin(3)=S^3, \quad Sp(1)^2=Spin(4)=S^3\times S^3,\quad Sp(2)=Spin(5)=S^3 (\times S^{7}.
\end{equation}
There are analogous results  for the  octonions, after $G_2$
stabilization.  The series goes up to dim $3$, but not beyond. This
is due to the  lack of associativity.
  We just write the results, adding the sphere exponents
\begin{eqnarray}
&G_2=SOct(1) \approx S^3 (\times S^{11}; \quad  Spin(8)=Oct(1)^2
\approx S^3 ( \times S^7 ( \times S^7 (\times S^{11} \\
&Spin(9):= Oct(2) \approx S^3 ( \times S^7 ( \times S^{11} (\times S^{15};\quad
F_4:=SOct(3) \approx S^3 ( \times S^{11} ( \times S^{15} (\times S^{23} \nonumber
\end{eqnarray}
where by the prefix "$S$" we mean the unimodular restriction (no $S^7$ factors).
This is similar to $SO$ and $SU$ for $\mathbf R$ and $\mathbf C$ respectively.
 The usefulness of the notation can be seen e.g. in the projective line and plane:
\begin{eqnarray}
&HP^1=S^4=Sp(2)/Sp(1)^2 \; \mbox {corresponds to} \; OP^1=S^8= Spin(9)/Spin(8), \\
&CP^2=S^5/S^1=SU(3)/SU(2) \; \mbox {corresponds to} \;
OP^2=SOct(3)/Oct(2)= F_4/Spin(9). \nonumber
\end{eqnarray}
The later is called the Moufang plane (Moufang 1933; to call it the Cayley plane is historically inaccurate).

In any case, this is just a notational convention, that we find useful, if carefully employed.
We finish by remarking that no use has been made so far of the fundamental {\it triality}
 property of the $O(8)$ group and the octonions, namely $Out[Spin(8)]\equiv Aut/Inner=S_3$,
 the order three symmetric group. Perhaps in a deeper analysis this triality will show up in particle physics.

{\bf Acknowledgments.}   AB  would like to thank  R. Ahl Laamara, B.
Belhorma, P.  Diaz,  L.B. Drissi, M. P. Garcia de  Moral,    J. Rasmussen, E.H. Saidi,A. Sebbar and M. B. Sedra for collaboration  on related subject.
He would also  thank  GNPHE and UFR-PHE-Rabat for hospitality (the end of 2007).
AS acknowledges discussions with P. Diaz, M.P. Garcia de Moral and E.H. Saidi.
This work has been supported by CICYT (grant FPA-2006-02315) and DGIID-DGA (grant 2007-E24/2), Spain. 
We thank also the support by  Fisica de altas energias: Particulas, Cuerdas y Cosmologia, A9335/07.

\end{document}